# Electrochemical tuning of capacitive response of graphene oxide

**Sanjin J. Gutić[1], Dževad Kozlica[1], Fehim Korać[1], Danica Bajuk-Bogdanović[2], Miodrag Mitrić[3], Vladimir M. Mirsky[4], Slavko V. Mentus[2,5], Igor A. Pašti[2]***

[1]University of Sarajevo, Faculty of Science, Department of Chemistry, Zmaja od Bosne 33-35, Sarajevo, Bosnia and Herzegovina

[2]University of Belgrade – Faculty of Physical Chemistry, Studentski trg 12-16, 11158 Belgrade, Serbia

[3]Vinča Institute of Nuclear Sciences, University of Belgrade, P.O. Box 522, 11001 Belgrade, Serbia

[4]Institute of Biotechnology, Department of Nanobiotechnology, Brandenburgische Technische Universität Cottbus-Senftenberg, 01968 Senftenberg, Germany

[5]Serbian Academy of Sciences and Arts, Knez Mihajlova 35, 11000 Belgrade, Serbia

---

[*] **corresponding author,** e-mail: igor@ffh.bg.ac.rs

**Abstract**

Increasing energy demands of modern society require deep understanding of the properties of energy storage materials as well as their performance tuning. We show that the capacitance of graphene oxide (GO) can be precisely tuned using a simple electrochemical reduction route. *In situ* resistance measurements, combined with cyclic voltammetry measurements and Raman spectroscopy, have shown that upon the reduction GO is irreversibly deoxygenated which is further accompanied with structural ordering and increasing of electrical conductivity. The capacitance is maximized when the concentration of oxygen functional groups is properly balanced with the conductivity. Any further reduction and de-oxygenation leads to the gradual loss of the capacitance. The observed trend is independent on the preparation route and on the exact chemical and structural properties of GO. It is proposed that an improvement of capacitive properties of any GO can be achieved by optimization of its reduction conditions.

# 1. Introduction

Among the most common routes for the preparation of graphene, which include chemical vapor deposition, epitaxial growth, micromechanical exfoliation and others, only (electro)chemical reduction of graphene oxide enables preparation of larger quantities of this material at reasonable expense. However, this cheap and scalable procedure generally leads to the *low-quality* graphene with a number of different non-reduced functionalities and structural defects.[1] Being considered as a drawback for some applications (like in electronics), functional groups and structural defects can be used as active sites essential for the applications of these materials in chemical sensors or in electrochemical systems.[1-3]

Zhou *et al*. were one of the first who reported an electrochemical reduction of graphene oxide (GO).[4] It did not take long until the importance of this approach for graphene production was recognized.[5] The advantages of this approach include a high efficiency of removal of oxygen groups, mild reaction conditions and an avoidance of toxic chemicals (such as hydrazine). Electrochemical reduction enables also a precise control of the reduction potential and reduction time, which is essential for the reproducibility of the process. In general, three different approaches for electrochemical reduction of GO can be distinguished. First, GO can be reduced as a film deposited on a conductive substrate[4,6-10] or it can be reduced from a suspension. In all the cases, an irreversible reduction of GO is observed, suggesting an efficient de-oxygenation of the material, which cannot be reverted upon exposure to high anodic potentials. Second, the reduction can be performed in aqueous[11,12] or in organic[12,13] suspensions. The electrochemical reduction of GO suspended in organic medium provides a better reproducibility and leads to the formation of higher-quality films[12], when compared to the products obtained from aqueous suspensions. Electrical conductivity of the suspension was found to be the critical parameter for the quality of the reduced GO film.[11] Electrochemical reduction of the deposited GO films can be performed by potential cycling. In this way, the reduction can be completed within only one potential cycle,[9,10] and, contrary to the GO reduction from suspensions, mass transfer limitations do not have significant impact on the reduction process. Namely, the reaction rate of the electrochemical GO reduction in GO suspensions is limited by diffusion, and the potential cycling has to be repeated many times in order to fully reduce the suspended material.[12] Cyclovoltammetric response of GO during this process depends on pH[4] and on the method of oxidation/exfoliation.[14] It was reported that prolonged sonication of GO suspension leads to a gradual decrease of the reduction current peak, while reduction potential remains constant.[15] The authors ascribed this phenomenon to the removal of highly oxidized domains in ultrasonic field.[15] In the recent spectroelectrochemical study it was suggested that the irreversible reduction of graphene oxide from aqueous suspensions depends on lateral size of graphene monolayer.[16] The same authors also found that graphene with a dominant OH content can be reduced easier than graphene with a high number of epoxy groups on basal plane.

The methods allowing one a controlled reduction were proved to be useful for precise modification of GO for applications in electrochemical systems.[17-20] These experimental results regarding the performance tuning have also been supported by computational studies.[21-24] For example, Ambrosi and Pumera have shown that the electron transfer properties of GO can be controlled precisely using electrochemical reduction.[18] If the reduction proceeds to lower cathodic potentials, the $\pi$ electron system is restored and the concentration of oxygen functional groups decays monotonically, while the electrocatalytic properties (probed using ferro/ferri-cyanide system) are improved.[18] Similar conclusions were derived by Liu *et al.*[25] who also demonstrated a reduced charge transfer resistance with the decrease of the GO reduction potential using impedance measurements. However, despite an intensive research, there is still no consensus regarding the impact of different structural and chemical parameters on the capacitive performance of graphene-based materials.[26-28] Some reports consider defects and functionalities to be more important than the specific surface area.[29] Namely, it was claimed that the contribution of the oxygen functional groups to the total capacitance of graphene-based materials can significantly surpass the contribution of the specific surface area.[26] On the other hand, some authors claim a negative impact of oxygen functionalities and the basal plane defects on the capacitive performance.[30] However, it is most likely that the overall capacitive response is governed by a fine balance between many structural, chemical and physical properties of the graphene-based materials. Recent theoretical study of the interactions of alkaline metals with oxidized graphene[31] suggested that an optimal concentration of oxygen functional groups should be found. This enhances the charge/metal ion storage capacity of graphene, while the material preserves high conductivity, which can be essential for electrochemical applications. Also, based on the computational modelling it was suggested that the aggregates of OH-groups on graphene basal plane are stable during electron transfer between alkali metal atom and oxidized graphene, thus suggesting that such aggregates should be able to store charge in the pseudocapacitive manner.[32] This is very important as the highly oxidized domains on graphene basal plane were found to preserve the high conductivity of graphene.[33] Indeed, the aggregation of oxygen functionalities by mild thermal treatment and consequent emergence of oxidized islands (sp$^3$-hybridized or graphene-oxide) on a highly conductive graphene surface was experimentally observed by Kumar *et al.*[33] Just recently, the importance of the highly oxidized domains for the capacitive performance was confirmed by Liu *et al.*[34] This results, in combination with previously mentioned studies, emphasizes the importance of both the concentration of oxygen functional groups and the conductivity of reduced GO for capacitive applications.

While analyzing charge storage properties of different graphene-based materials, we reported that the total capacitance of electrochemically reduced thin GO film, containing a conductive component (carbon powder Vulcan XC-72), depends on the depth of cathodic polarization.[10] Namely, when the reduction potential was increased in the cathodic direction, capacitive response passed through a maximum and then decayed, for the GO reduced at very low potentials.[10] The observed behavior was explained in terms of the pseudocapacitive contributions of the oxygen functionalities and the conductivity of the material, which was a point of speculation at that moment. It was suggested that there is an optimal ratio between the concentration of oxygen functional groups and the conductivity, leading to the maximal capacitance.[10] Here we show that such behavior is a *general* feature of graphene oxide. Using direct *in situ* measurements of the conductivity of thin electrochemically reduced GO films in a combination with electrochemical and structural characterization, we demonstrate a universal behavior of electrochemically reduced GO films in terms of tuning of its capacitive response.

**2. Experimental**

*2.1. Materials*

The commercial GO sample, denoted as GO-ACS, was described in details in our previous work.[10] Additionally, five home-made GO samples were characterized and used in this work. The sample denoted hereafter as GO-Ec was obtained by electrochemical oxidation/exfoliation of graphite (obtained from commercial zinc-carbon battery) by adopting the procedure descried in Ref. [35]. The pH-neutral solution and a two-compartment electrolytic cell were used. Other four home-made GO samples were prepared by the improved modified Hummers method.[36] Depending on the pretreatment of the natural graphite precursor and the selection of the particle size of the precursor, the samples are denoted as follows: (i) GO-0 for the material obtained without any graphite pretreatment; (ii) GO-50, obtained from graphite particles smaller than 50 μm; (iii) GO-40, obtained from graphite particles smaller than 40 μm; and (iv) GO-40HF for the material obtained from graphite particles smaller than 40 μm treated in hot concentrated aqueous HF for 6 hours. Upon the synthesis and purification, the dispersions of home-made samples were diluted to provide the absorbance at 380 nm identical to that of the aqueous GO-ACS dispersions having the concentration 0.9 mg cm$^{-3}$. The prepared dispersions remained stable over time and were used as such.

*2.2. Characterization*

The X-ray Powder Diffraction (XRPD) patterns of investigated samples were obtained on a Philips PW-1050 diffractometer, operated at 40 kV and 20 mA, using Bragg–Brentano focusing geometry and Ni-filtered Cu K$\alpha_{1,2}$ radiation. The patterns were taken in the 6–80° $2\theta$ range with the step of 0.05° and exposure time of 4 s per step.

Raman spectra, excited with diode-pumped solid state high brightness laser (excitation wavelength 532 nm), were collected on a DXR Raman microscope (Thermo Scientific, USA) equipped with an Olympus optical microscope and a CCD detector. The laser beam was focused on the sample using objective magnification 10×. The scattered light was analyzed by the spectrograph with a 900 lines mm$^{-1}$ grating. Laser power on the sample was kept at 2 mW to prevent thermal degradation of the samples.

Attenuated total reflectance Fourier transform infrared spectroscopy (ATR-FTIR) has been used to characterize investigated samples in the 400−4000 cm$^{-1}$ range with a spectral resolution of 4 cm$^{-1}$ at room temperature by Thermo Scientific Nicolet iS50 FT-IR spectrometer equipped with built-in all reflective ATR diamond.

Scanning electron microscopy (SEM) combined with energy-dispersive X-ray spectroscopy (EDX) was performed using Phenom Pro-X (Holland) electron microscope.

*2.3. Electrochemical experiments*

Electrochemical measurements of all the materials, including the reduction procedures, were performed in a standard three-electrode electrochemical cell, with Ag/AgCl (KCl$_{sat}$) as the reference and a large Pt foil as the counter electrode. Unless otherwise stated, the electrode potentials reported here are referred to Ag/AgCl (KCl$_{sat'd}$) electrode. Experiments were performed with potentiostat/galvanostat PAR 263A, controlled by PowerCV and PowerSTEP interfaces. Electrochemical cell was extensively purged with argon (99.999%) before and during all the experiments. All electrochemical experiments were performed in de-aerated 1 mol dm$^{-1}$ aqueous K$_2$SO$_4$.

Working electrode was prepared by direct drop-casting of GO dispersion on a glassy carbon disc with the diameter of 5 mm, except for the preparation of GO films for Raman measurements. In the later case thoroughly cleaned Cu discs were used as the substrates. Ethanol/water (4/6 v/v) dispersions of GO samples, containing 0.05 % of Nafion, were drop-casted on the GC surface and dried under vacuum. As we are interested in the overall trends of the capacitive response, upon the reduction of GO and the described drop-casting procedure induces relative variations of loaded GO within 7%, a new GO electrode was prepared for each reduction potential (relative uncertainties of drop-casting method were estimated using the series of nine consecutive measurements). Hence, the capacitance increment factors reported here were determined for each electrode after reduction at a given potential as a ratio between the capacitance after and before reduction. After drop casting of GO onto the GC surface and vacuum evaporation of solvent, GO film was cycled for ten time (potential sweep rate 10 mV s$^{-1}$) between 0.80 and −0.50 V. This resulted in a stable voltammetric response. Then the capacitive voltammograms were recorded in the same potential window (between 0.80 and −0.50 V) before and after the reduction at one of the indicated potentials. Finally, the capacitance increment factors were determined as described above. Reduction of graphene oxide films was performed potentiostatically at different potentials during 10 seconds time intervals. Using chronoamperometric measurements, this time interval was found to be sufficiently long to complete GO reduction at each given potential. If not specifically stated, the results reported here refer to the measurement at the potential scan rate of 100 mV s$^{-1}$.

The *in situ* electrical resistance measurements were performed on the system described in Refs [[37,38]]. The measurements were performed for GO-ACS sample. Thin film was deposited on the interdigitated gold electrode on a glass substrate. The 50 mV pulses of alternating polarity were applied to the outer electrodes of the interdigitated structure, the resulting current was measured, and the resistance was calculated. The measurement system has the sampling interval of around three seconds. Potential of the working electrode was controlled with respect to the reference electrode (Ag/AgCl (KCl$_{sat'd}$) electrode) and changed according to a staircase program between −0.6 V and −1.6 V and back, with a step of 0.1 V. The electrode was held at each potential for two minutes.

## 3. Results and discussion

Cyclic voltammetry, performed with the GO-ACS sample (Fig. 1, top panel), clearly demonstrate an irreversible nature of the GO reduction process. Presented cyclic voltammograms are performed with a single sample, extending cathodic vertex potential in each successive scan. In neutral solution, according to the results presented in Fig. 1, the reduction commences at the potentials slightly below −0.8 V *vs*. Ag/AgCl. Upon extending cathodic vertex potential, GO is reduced to higher extent, while no reverse oxidation peaks in anodic scan can be identified. The reduction is practically complete upon extending cathodic vertex potential to −1.2 V and no significant faradaic response is seen upon further extension of reduction potential in subsequent potentiodynamic scans (Fig. 1). From the repetitive cyclic voltammograms of GO-ACS reduction it can also be concluded that the electrochemical reduction is very fast process and that, for a given cathodic vertex potential, most of the reducible oxygen groups are removed from the GO surface in the first scan.

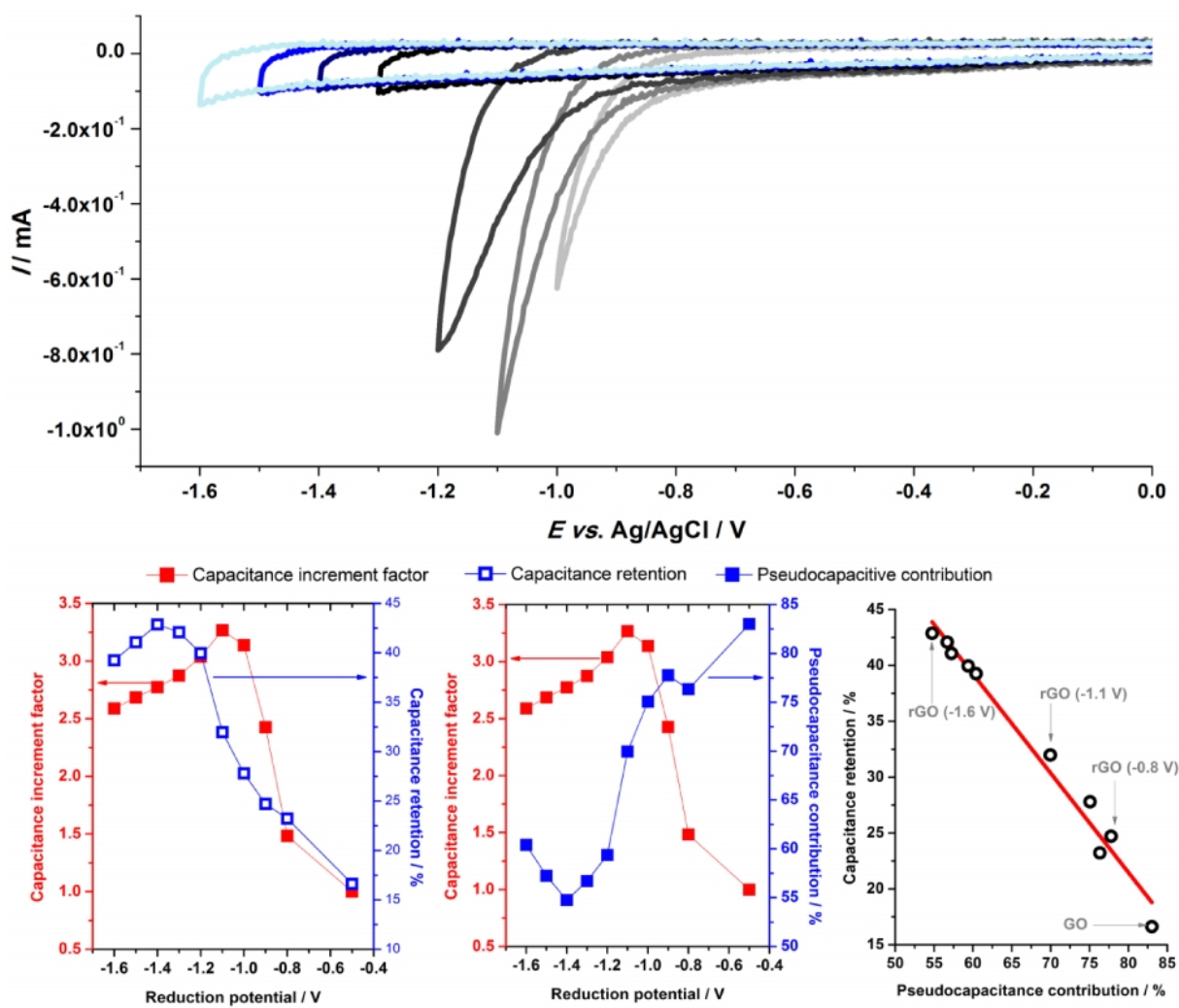

**Figure 1.** Successive cyclic voltammograms with negative shift of cathodic vertex potential (upper panel) and the dependence of the capacitance increment factor, capacitance retention and pseudocapacitance contribution on the GO reduction potential (lower panel). Linear correlation between the capacitance retention and pseudocapacitance contribution for electrochemically reduced GO-ACS at different potentials is presented. Capacitance of reduced GO-ACS samples was investigated in the potential window between 0.80 and −0.50 V vs. Ag/AgCl.

Following the work of Ambrosi and Pumera,[18] such GO reduction results in an increase of the C/O ratio from ~3, for starting GO, to ~10 for electrochemically reduced GO. Considering the accompanying changes in the capacitive response of the reduced GO film, one can see that the capacitance increases by 50% already for GO reduced at −0.8 V at which a very small fraction of oxygen functional groups is removed. After the initial reduction steps at mild potentials, the capacitance increases significantly, achieves maximum for GO-ACS reduced at −1.1 V, and decreases upon further negative shift of the reduction potential (Fig. 1, lower panel). It is important to emphasize that we report normalized capacitances of reduced materials with respect to the capacitance of the starting GO material (i.e. capacitance increment factors), instead of reporting gravimetric capacitances. This is necessary because considerable amount of oxygen is removed upon reduction, leading to a significant decrease of the mass of the material on the electrode. Considering that the mass of removed oxygen cannot be precisely determined, and that some carbon containing species can also be removed, normalization of the measured capacitance(s) to the mass of starting GO could give significantly lower values of the gravimetric capacitance. In overall, capacitance increment factor of 3.3 is reached when GO-ACS is reduced at −1.1 V. Previously, Zhang *et al*. reported an increase of GO capacitance after electrochemical reduction by a factor of ~15, however, no connection of the capacitance increase with the reduction potential was shown.[39]

Following reduction of GO-ACS film, we observed the changes in Raman spectra, showing strong dependence on the potential of reduction (Fig. 2). While Raman spectra of as-received GO-ACS shows typical features of a highly disordered material,[40-42] the reduction of the GO-ACS film leads to the evolution of the Raman spectra. In particular, we see the change in the relative intensities of D (disorder) and G (graphitic) band, located around 1350 and 1580 cm$^{-1}$, respectively. Pronounced changes in relative intensities of $I_D$ and $I_G$ bands are clearly seen upon reduction at −1 V (Fig. 2). The calculated $I_D$/$I_G$ ratios for all the films are reported in Fig. 3. This clearly indicates the change of the structural order of reduced GO-ACS. An increase of the $I_D$/$I_G$ ratio up to the reduction potential of −1.2 V is noticed. Further shift of the reduction potential to lower cathodic potentials leads to the decay of the $I_D$/$I_G$ ratio. As the reduction potential decreases, the D-band becomes more pronounced while the D'-band appears as the shoulder of the G-band located around 1580 cm$^{-1}$. Moreover, the overtone bands, i.e. 2D region,[42] also start to develop as indicated in Fig. 2 for the samples reduced at potentials below −1.0 V. These bands are found at around 2670 cm$^{-1}$ and 2930 cm$^{-1}$. Following Ref. [[42]] the first band is denoted as 2D (G') while the second one is denoted either as D+D' or D+G. The appearance of the 2D region, and its evolution from bump-like shape, is typical for GO reduction process.[42] In particular, these bands are consider to be defect-activated and dependent on the number of layers in graphite/graphene. Moreover, their appearance is also associated with wrinkled graphene layers which contain significant amount of defects.[42] As we do not expect that GO layers re-stacking can take place when deposited in the form of a thin film on the electrode surface, due to the addition of Nafion binder (Section 2.3), the evolution of 2D region and the appearance of D' band is due to the defects formed upon electrochemical reduction. This can be ascribed unambiguously to the removal of oxygen groups along with a certain fraction of carbon atoms, as mentioned previously.

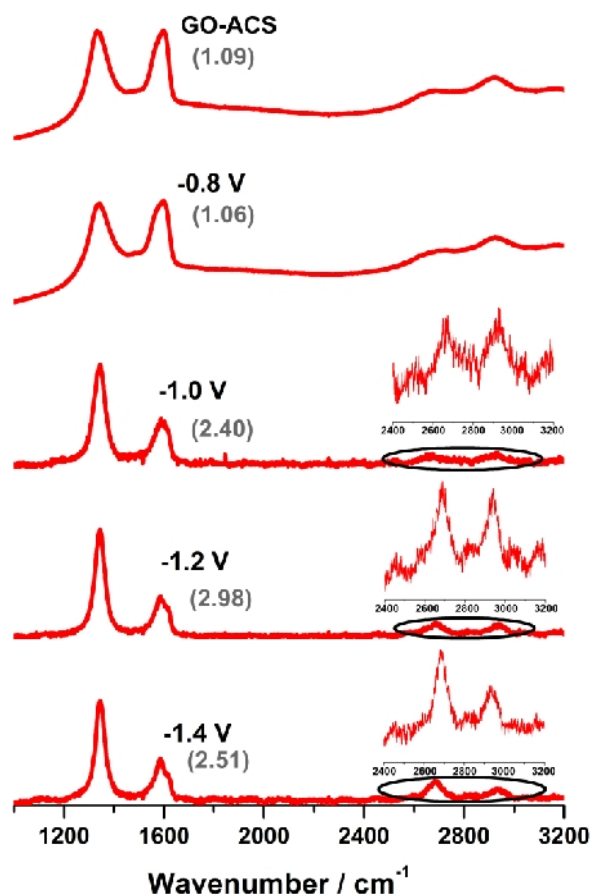

**Figure 2**. Raman spectra of thin GO-ACS film reduced at different potentials (as indicated in the figure). Values in parentheses give calculated $I_D/I_G$ ratios. Insets show emphasized spectral region of the overtone bands.

The significant changes of the Raman spectra and the increase of the $I_D/I_G$ ratio coincide with the onset potential for GO reduction, suggesting that de-oxygenation of GO-ACS is accompanied with significant changes of the disorder level. According to the work of Lucchese et al.[41] it is suggested that the changes of the $I_D/I_G$ ratio point to structural ordering during the reduction process and restoration of the π electron system, which, in overall, goes beyond the formation of structural defects due to removal of oxygen functional groups, evidences in the evolution of the 2D band region, as discussed previously. However, we expect that the structural ordering of GO-ACS films reduced at very low potentials is slightly lower compared to the films reduced at potentials around −1.2 V,[41] where the capacitance increment factor already starts to decay (Fig. 1). Previously, almost negligible changes of $I_D/I_G$ ratio upon electrochemical reduction of GO were reported.[43] However, just recently, it was reported that the initial ordering of GO upon thermal reduction is followed by disordering when reduction takes place at high temperatures.[44]

Other authors also observed an increase of the electrochemical response of GO upon intensification of the reduction conditions[20,44] as well as upon the increase of the reduction time.[17,27] The observed trends were discussed in terms of the removal and/or evolution of oxygen functionalities in all the cases. According to Toh *et al.*,[20] even after reduction at −1.5 V a considerable amount of epoxide and hydroxyl groups remains at the graphene plane. Furthermore, same authors found only slight decrease of the oxygen content upon reduction between −1.2 and −1.5 V, while in the same time, using XPS analysis, they found considerable increase of the concentration of OH groups.[20] However, Ambrosi and Pumera found that carboxylic groups can be reduced only at very high cathodic potentials, while carbonyl and epoxy groups are effectively reduced already at very low cathodic potentials.[18] While there is no consensus on the evolution of the oxygen functional groups upon the electrochemical reduction, the observed increase of the fraction of $sp^2$ hybridized carbon atoms leads to the conclusion that the de-oxygenation restores the $\pi$ electron system.[18,20] A natural question is: whether the changes of the capacitance can be related *solely* to the evolution of the oxygen functional groups upon the reduction?

In order to answer this question we first compare the capacitance increment factors for GO-ACS presented here and the ones previously reported for the GO-ACS films containing conductive component.[10] In the latter case we observed much smaller capacitance increment factors (increase up to ~ 50% of the initial value before reduction). Hence, relative increase of the capacitance in the absence of conductive additive is noticeably higher for all the reduction potentials. We suggest that the conductive component in the GO film, used in Ref. [[10]], reduces the impact of the restoration of the $\pi$-electron system on the conductivity, and, consequently, the capacitance increase observed at milder reduction potentials. Moreover, the question is how reduction of GO-ACS at mild potential (−0.8 V) can increase the capacitance by 50% (Fig. 1) when only 6% of reducible O-functional groups is removed at this potential (as estimated from the integrated CV of GO-ACS reduction). In order to obtain an insight into the effect of the electrical conductivity, we performed *in-situ* measurements of the electrical resistance during the reduction of GO-ACS at different potentials (Fig. 3).

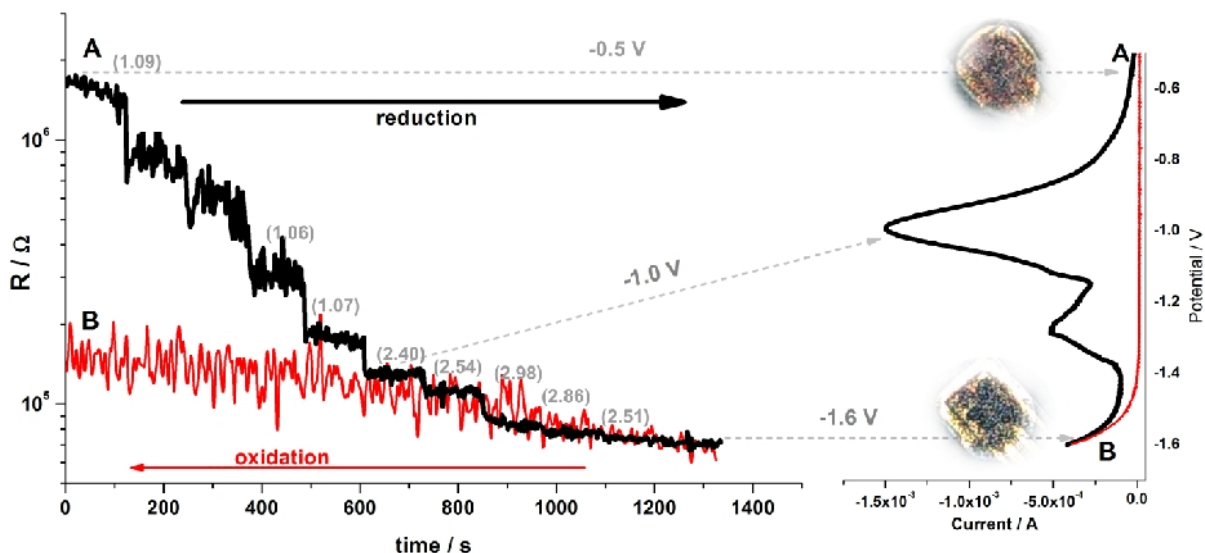

**Figure 3**. Left: the dependence of the *in situ* measured electrical resistance of the GO-ACS film during reduction (thick line black) and oxidation (thin line red). Electrical resistance (left) is associated with the cyclic voltammogram of the GO-ACS reduction (right). The numbers in parentheses give the determined $I_D/I_G$ ratios for each reduction potential. Insets show the photographs of GO-ACS films before (A) and after the reduction (B).

We see that very small structural changes of GO-ACS reduced at −0.8 V, as indicated by small variations of the $I_D/I_G$ ratio (Figs. 2 and 3), are followed by an order of magnitude drop in the resistance of GO film. At −1.0 V, where roughly 60% of oxygen groups is reduced, the $I_D/I_G$ ratio increases sharply and the resistance continues to decay, showing rather small change upon further lowering of reduction potential (Fig. 3). In the reverse scan a gradual increase of the resistance is observed. However, this is also the case for the fully reduced carbon. Namely, using reduced carbon samples, we observed the variations of the electrical resistance during potential sweep between deep anodic and deep cathodic potentials. The resistance passed a small maximum (increase up to 70%) at intermediate potentials (not shown here). Irreversible changes of GO are also visually clear as the color of the GO film changes from brown to black upon reduction at deep cathodic potentials (Fig. 3, inset photographs). It should be noted that the measured resistance contains both bulk and contact resistance of the reduced GO film. Hence, the absolute values of the measured resistance depend on both factors. Nevertheless, the relative changes of the resistance agree with earlier reports on the resistance changes of oxidized graphene upon annealing and chemical reduction.[45]

Based on the obtained results we can explain the trends in the capacitive response of reduced GO by two factors: (i) a decrease of the amount of oxygen functional groups and (ii) a decrease of the electrical resistance. At mild reduction potentials (positive to the CV reduction peak) a relatively low fraction of oxygen functional groups is removed, and the capacitance increases due to significant increase of the electrical conductivity of the film. In this case initially low conductivity of the as-obtained sample hinders capacitive response. Hence, the redox activity of remaining oxygen functional groups is enabled *via* improved electron transport through the reduced GO sheets. Upon reaching maximum capacitance, a decrease of the capacitive response is observed while the resistance remains nearly constant. This is caused by lowering the fraction of the redox active groups on the surface upon extensive reduction.

Besides the possibility to tune capacitive response of electrochemically reduced GO, we analyzed the contributions of pseudocapacitance and double layer capacitance to the total capacitance of the reduced graphene oxide. This was done by the extrapolation of the charges measured during potential sweeps at various rates to infinitely slow ($v \rightarrow 0$ mV s$^{-1}$) and infinitely fast ($v \rightarrow \infty$ mV s$^{-1}$) scan rates, as proposed by Lee *et al*.[46] The amount of charge obtained from the first extrapolation relates to the total capacitance, while the amount of charge from the second extrapolation relates to the double layer capacitance. Their difference corresponds to the pseudocapacitance of the material. It is important to emphasize that the values obtained by this method should be taken with caution. In our interpretation, "pseudocapacitive contribution" includes all the possible processes that can limit charge/discharge rate, including faradaic processes and the diffusion in micropores and the interparticle voids. As can be seen from Fig. 1, the steepest decrease in "pseudocapacitive contribution" is observed in the potential range where the capacitance passes through maximum. Also, it occurs at the potential range in which the highest conductivity increase is observed (Fig 3.). Therefore, a decrease in "pseudocapacitive contribution" only partially correlates with an increase in capacitance: if the reduction is performed at more negative potentials than −1.1 V, the decrease of both quantities is observed. This is due to the removal of oxygen functional groups which contribute charge storage, while the increase of the conductivity cannot compensate for these losses. A small increase of the "pseudocapacitive contribution" upon reduction at very negative potentials (Fig. 1) can be ascribed to the increase of disorder level observed by Raman spectroscopy.

In addition, the capacitance retention was calculated as the ratio between the capacitance measured at 800 mV s$^{-1}$ and the capacitance extrapolated to $v \rightarrow 0$ mV s$^{-1}$. We observed that the capacitance retention decreased linearly with the pseudocapacitive contribution (Fig. 1). We have previously found this type of behavior in different aqueous electrolytes for other graphene-based materials as well.[10] However, the highest capacitance retention is not correlated with the maximum increase in the capacitance. Instead, it was observed for GO reduced at deep cathodic potentials (Fig. 1), which is in agreement with the results of other authors as well.[26] It is important to emphasize that the observed trends in the capacitance retention could possibly explain the discrepancies in the reports of different authors regarding the contribution of the oxygen functional groups to the capacitance of carbon materials. While the capacitance values obtained at high scan rates (or high charge/discharge currents) point towards negative role of oxygen functionalities, the measurements at low scan rates (or low charge/discharge currents) indicate beneficial role of the oxygen functionalities. It is important to emphasize that the trends in capacitance increment factors discussed here, as well as in our previous report,[10] are fully independent on the applied scan rates.

The same trends in capacitive response were observed for the series of home-made samples. GO-Ec (Figs. 4 and 5) shows irreversible reduction at somewhat more negative potentials, when compared to GO-ACS, accompanied with a much less pronounced changes of Raman spectra, in line with some earlier reports[43] (Fig. 4). Calculated $I_D/I_G$ ratios are provided in Fig. 4.

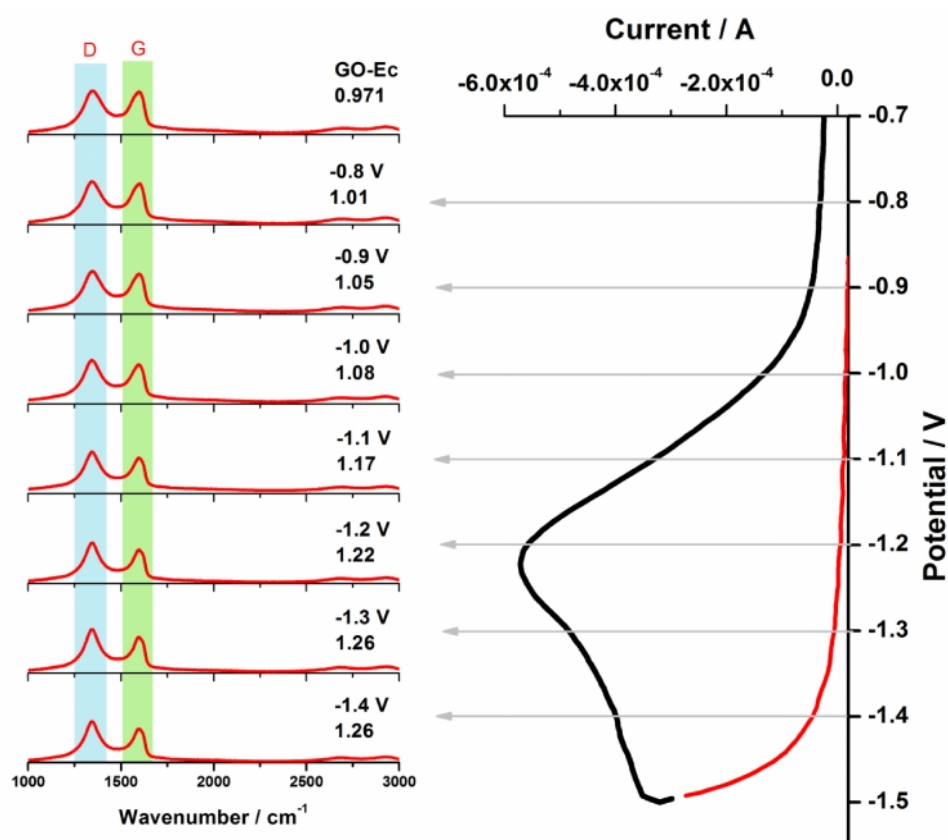

**Figure 4**. Evolution of the Raman spectra of the GO-Ec sample with irreversible reduction at low cathodic potentials. The numbers below reduction potentials (which are negative) indicate the calculated $I_D/I_G$ ratio. Cyclic voltammogram on the right shows the reduction of GO-Ec (the thick line corresponds to the direct scan in cathodic direction, the thin line corresponds to the reverse scan in anodic direction).

However, the XRD pattern (Fig. 5) of the same sample shows a very distinct (002) diffraction peak of graphite[47,48] suggesting the existence of graphitic fraction in the GO-Ec sample. In contrast, Raman (Fig. 4) and ATR-FITR spectra (Fig. 5) show significant structural disorder and confirm the existence of oxygen functional groups on the sample surface. The typical features of GOs with C-O (ether) and phenol (C-OH) peaks at 1064, 1222 (C-O) and 1360 cm$^{-1}$ (C-OH) and carboxyilic C=O stretching vibrations at 1620 and 1736 cm$^{-1}$ are seen in the ATR-FTIR spectrum of GO-Ec[49]. Peak at 1600 cm$^{-1}$ is also contributed with C=C vibration mode which is centered at this wavenumber. Also, a broad absorption peak at ca. 3200 cm$^{-1}$ and peaks at 2800 and 3600 cm$^{-1}$ indicate the presence of considerable amount of OH groups. The peak at 2800 cm$^{-1}$ could also be contributed to C-H vibrations. In order to exclude possible incorporation of other elements into GO-Ec during the synthesis we performed EDX analysis of the sample deposited on Cu substrate and confirmed that only C and O are in the sample (Fig. 5). Oxygen concentration in the sample was found to be 17 at.%.

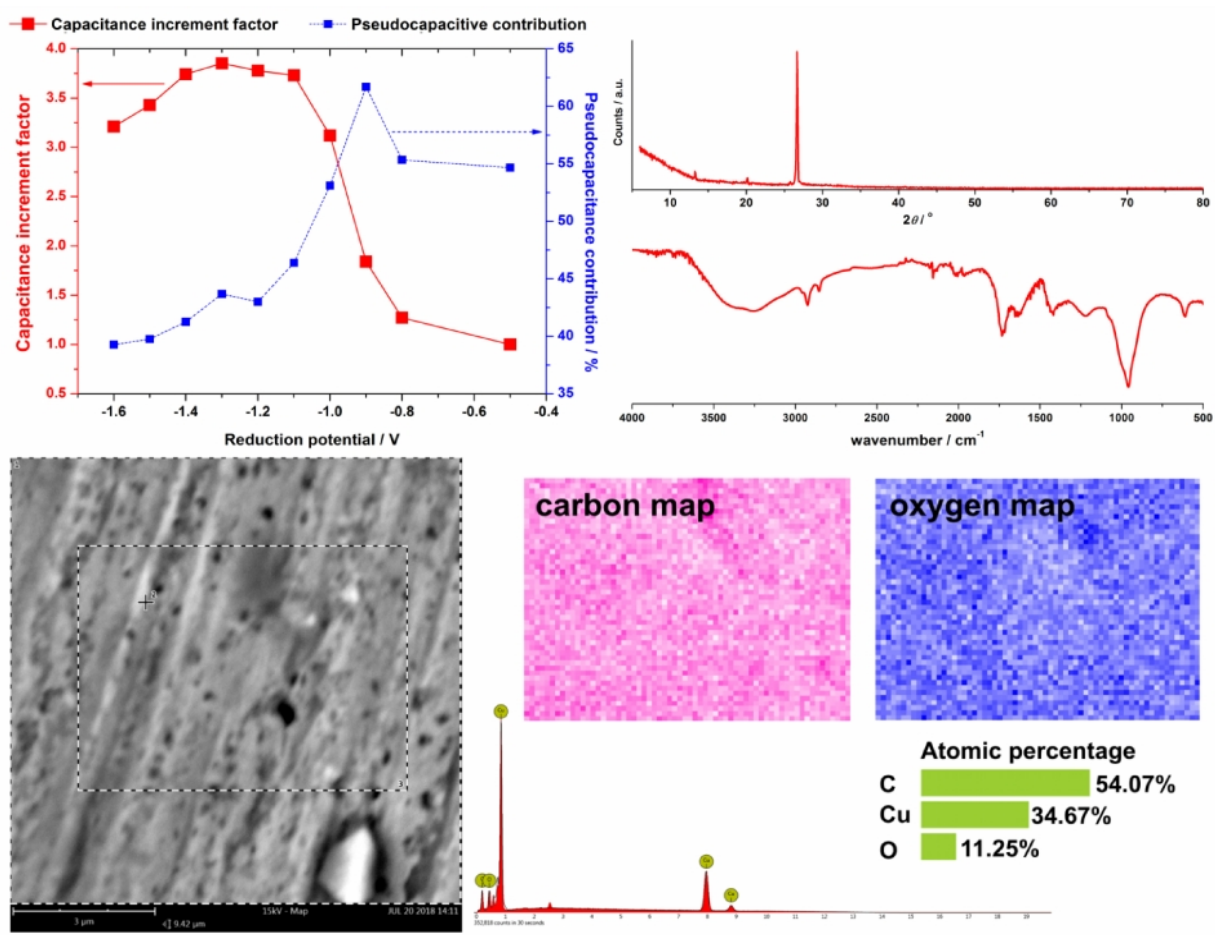

**Figure 5.** Top lLeft: Dependence of capacitance increment factor and pseudocapacitance contribution on the reduction potentials; top right: XRD pattern and FTIR spectra of prepared GO-Ec. Pseudocapacitive contribution was calculated as explained for the GO-ACS sample. Bottom: SEM image of GO-Ec thin film on copper substrate with EDX analysis and oxygen and carbon elemental maps (mapped region is indicated in SEM micrograph).

Electrochemical behavior of GO-Ec matches that of GO-ACS. Upon the progression of GO-Ec reduction, the capacitance increases by a factor of ~3.7. The reduction potential which corresponds to the maximum capacitance increment factor coincides with the peak potential of the irreversible GO-Ec reduction (around −1.2 V). Considering the "pseudocapacitive contribution" (Fig. 5), we observe small increase followed by a monotonous decrease upon progressive reduction of GO-Ec. Again, this can be ascribed to removal of oxygen functional groups which contribute charge storage properties of the material.

Finally, we studied four different chemically produced GO samples (denoted as GO-0, GO-50, GO-40 and GO-40HF, Section 2.1). In contrast to GO-Ec these materials show typical XRD pattern of graphene oxide with reflections between $2\theta$ = 10.65° and 10.9°. This is an indication of an effective separation of graphene layers, with the distances between them being larger than 8 Å[48,50] (Fig. 6). There is also a weak diffuse graphitic (002) peak at $2\theta \approx 28°$, suggesting a small fraction of the remaining graphitic structures. ATR-FTIR spectra of these materials are characteristic for graphene oxides[49] while Raman spectra suggest a significant structural disorder (Fig. 6). This is expected because of the applied oxidation method[51] and the selection of the starting material, i.e. natural graphite.[52] According to the presented results, the four chemically prepared GOs have rather similar structural properties. However, their electrochemistry is marked with some significant differences.

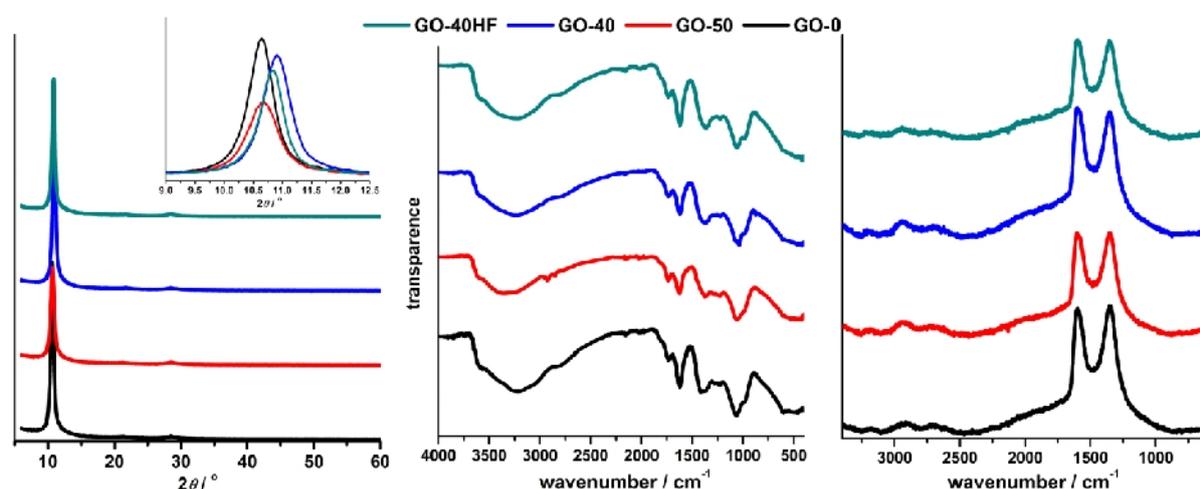

**Figure 6.** XRD patterns (left; inset shows typical GO reflections between $2\theta$ = 10.65° and 10.9°), ATR-FTIR spectra (center) and Raman spectra (right) of the four chemically prepared GO samples.

Cyclic voltammograms recorded with the thin films of chemically prepared GOs on glassy carbon (Fig. 7, upper panel) indicate a step-wise reduction process with three current peaks that arise from irreversible reduction of GOs. However, the most negative peak (at −1.7 V) becomes less pronounced in the following order: GO-0 > GO-50 > GO-40 > GO-40HF. In the same order the reduction peaks gradually overlap and shift to more negative potentials. It should be noted that the dependence of the electrochemical behavior of rGO on the precursor (i.e. graphite) particle size was previously observed by Tran *et al.*[53,54] In these reports, significant improvement of the electrochemical properties of rGO was observed upon decreasing graphite particle size, and it was discussed in terms of increasing oxidation efficiency upon decreasing grain size. In addition, improvement of electrochemical properties of (r)GO by HF treatment can be ascribed to the purification[55,56] and etching properties of HF,[57-59] which could further increase graphite oxidation and exfoliation efficiency.

Just like in the cases of GO-ACS and GO-Ec, upon the reduction the capacitive response increases and reaches maximum, but at different reduction potentials depending on the GO sample. A striking change of the capacitive response upon the reduction at different potentials is clearly visible in the change of the cyclic voltammograms in Fig. 7 (bottom left) for the GO-40HF sample, for which maximum capacitance is observed upon the reduction at −1.4 V. It can be seen that capacitive response changes from basically none to the characteristic rectangular shape cyclic voltammogram with a pseudocapacitance contribution. The pseudocapacitance is seen as a wide hump on the cyclic voltammograms in the potential range between −0.5 and 0.1 V. The potential of the maximum capacitance correlates with the potential of the first reduction peak on the cyclic voltammograms of the irreversible GO reduction (indicated by arrows in Fig. 7, upper panel). As the potential which corresponds to the maximum capacitance shifts from −1.2 V for GO-0 and GO-50 to −1.3 V for GO-40 and −1.4 V for GO-40HF, the maximum values of the capacitance increment factor increase in the same order: 40, 70, 90 and 110 for GO-0, GO-50, GO-40 and GO-40HF, respectively.

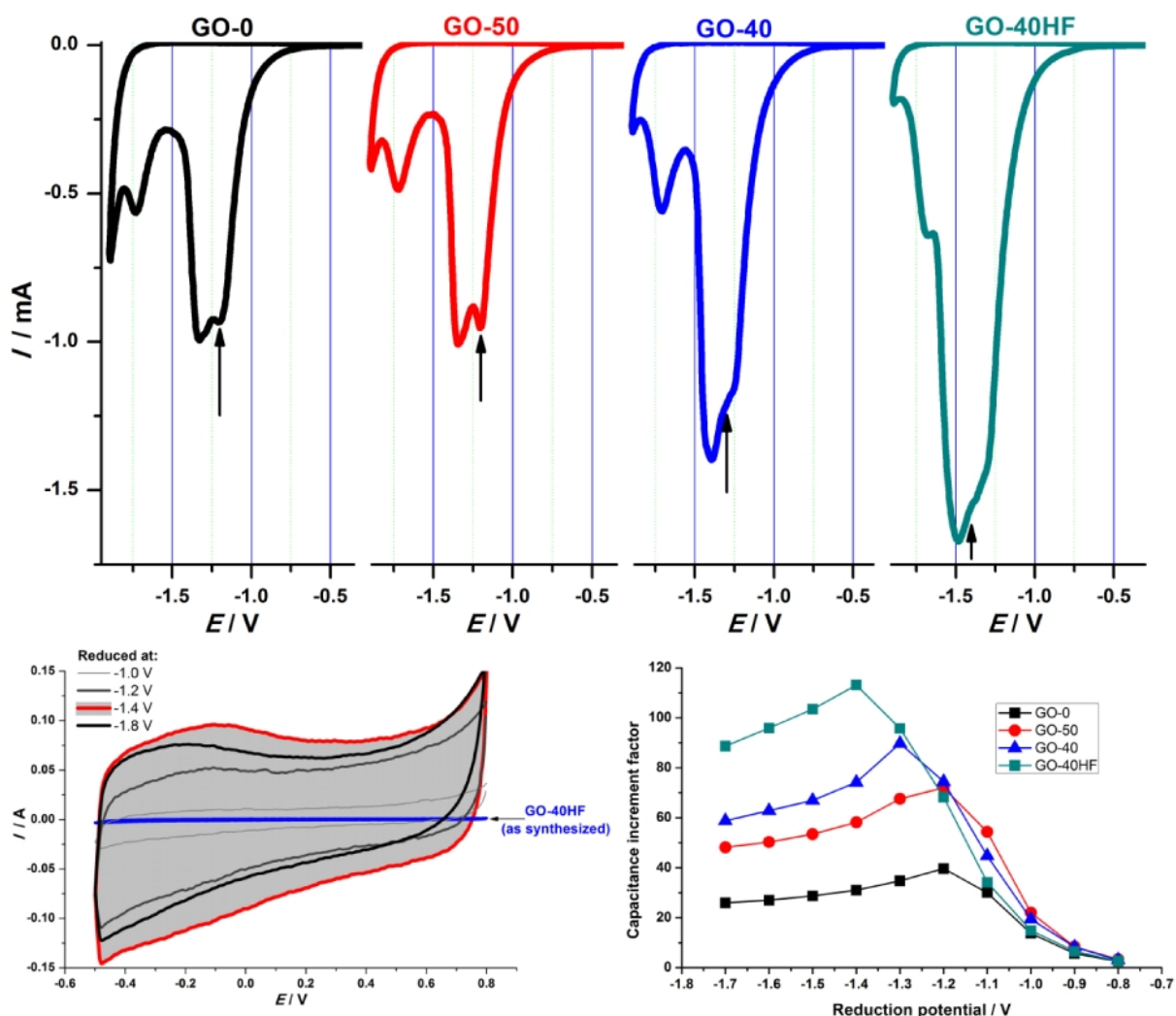

**Figure 7**. Upper panel: cyclic voltammograms of irreversible GO reduction. Arrows indicate the reduction potential at which maximum capacitance is reached. Bottom left: cyclic voltammograms of GO-40FH (as-prepared and upon reduction at different potentials). Shaded cyclic voltammogram gives the highest capacitive response and corresponds to GO-40HF reduced at −1.4 V *vs*. Ag/AgCl. Bottom right: dependence of capacitance increment factor on the reduction potentials for four home-made GO samples.

It can be observed that capacitance increment factors observed in this work vary to a significant extent, between 3.3 (GO-ACS) and 110 (GO-40HF). Such a large difference can be ascribed to many different factors such different oxidation levels of initial GO material, particle sizes, exfoliation degrees, conductivity of starting GO material and others. Definite conclusion on the exact role of each mentioned property is difficult to derive and beyond the scope of present work. However, we emphasize a general behavior of all GO samples, found in the maximized capacitance when proper reduction potential is found. Also, it is important to revisit previously discussed role of conductive component on the capacitance increment factors.[10] Smaller increment factor does not imply smaller specific capacitance - it is actually an indication of the positive effect of the conductive additive on the capacitance. In this case, smaller increment factor is the consequence of a higher initial capacitance of the starting non-reduced GO. As the conductivity of the reduced forms is high, the contribution of the conductive additive diminishes, leading to similar values of specific capacitance of electrochemically reduced GO with and without conductive additive. This, however, does not limit the application of present findings. First, conductive component is usually added to materials for electrochemical capacitors but electrochemical treatment is not combined with the addition of conductive component, and we previously confirmed that capacitance increases in the present of conductive component as well.[10] Second, present results indicate that the addition of conductive component can be avoided, at least in some cases, which simplifies processing of the electrode materials and the electrode preparation process.

## 4. Conclusions

We demonstrate that it is possible to tune finely the capacitive response of graphene oxide by electrochemical reduction under potentiostatic conditions. Upon the exposure to negative potentials graphene oxide undergoes irreversible changes which include de-oxygenation and structural ordering resulting in the restoration of the $\pi$ electron system and the conductivity increase. The capacitance of reduced graphene oxide achieves its maximum when the concentration of oxygen functional groups and the conductivity are properly balanced. This trend in the capacitance response may be interpreted in the following way. At early stages of the reduction progress increase of the capacitive response correlates with the conductivity increase, due to an improved current collection. However, upon further reduction at more negative potentials, oxygen functional groups, responsible for high pseudocapacitance, are depleted, which is accompanied by an overall capacitance decrease. The existence of an optimal reduction potential which maximizes the capacitance does not depend on the preparation method and the nature of the oxygen functional groups present on the surface. Hence, we conclude that an optimal reduction conditions can be found for any graphene oxide in order to maximize its capacitance. Also, based on the obtained results, it can be expected that the capacitance of graphene might be increased by controllable oxidation so that the conductivity is not affected while sufficient concentration of oxygen functional groups is introduced in the material. However, such a capacitance tuning in the "oxidation direction" might be much more difficult to control than the capacitance tuning in "reduction direction", like the one demonstrated here. In conclusion, we believe that the present work provides an insight into the tailoring of the capacitive properties of graphene-based materials and offers new strategies for the development of graphene-based electrochemical capacitors. Considering that the electrochemical methods are rather scalable, electrochemical tuning of the capacitance of graphene oxide has a high potential for a large scale application in the development of graphene-based electrochemical capacitors.

**Conflicts of interest**

There are no conflicts to declare.

**Acknowledgements**

S.V.M. and I.A.P acknowledge the support of the the Serbian Ministry of Education, Science, and Technological Development through the project no. III45014. S. V. M is indebted to Serbian Academu of Sciences and arts for funding this study through the project "Electrocatalysis in the contemporary processes of energy conversion". Authors are grateful to Prof. Kurt Kalcher from Karl-Franzens-Universität, Graz, Austria and Prof. Emir Turkušić from Faculty of Science, Univeristy of Sarajevo for providing us the GO-ACS sample.